\documentclass[twocolumn]{aa}
\usepackage{graphicx}
\begin{document}

   \title{The detectability of extrasolar planet surroundings}

   \subtitle{I. Reflected-light photometry of unresolved rings}

   \author{L. Arnold \inst{1}
          \and
          J. Schneider\inst{2}
          }

   \offprints{L. Arnold}

   \institute{Observatoire de Haute-Provence CNRS 04870 Saint-Michel-l'Observatoire, France\\
             \email{arnold@obs-hp.fr}
         \and
             Observatoire de Paris-Meudon 92195 Meudon Cedex, France\\
             \email{Jean.Schneider@obspm.fr}
             }

   \date{Received November 20, 2003; accepted March 9, 2004}

   \abstract{
   It is expected that the next generation of high-contrast imaging instruments will deliver
   the first unresolved image of an extrasolar planet. The emitted thermal infrared light from the planet
   should show no phase effect assuming the planet is in thermal
   equilibrium. But the reflected visible light will vary with phase angle.
   Here we study the photometric variation of the reflected light with orbital phase of a ringed
   extrasolar planet. We show that a ring around an extrasolar planet, both obviously unresolved, can be detected by its specific
   photometric signature.
   A simple quantitative model is discussed, taking into account the basic optical and geometrical properties of the ringed planet.
   }

   \maketitle
%
\section{Introduction}
The discovery of extrasolar planets by radial velocity measurements has provided the first dynamical characteristics of planets (orbital
elements and mass). The next step will be to investigate physical characteristics (albedo, temperature, radius etc.) and planet
surroundings. Among the latter are the satellites and rings of planets. Here we present the first of a series of papers on the
detectability of exoplanet surroundings, namely rings and satellites. It would seem {\it a priori} that their detection will
require imaging at very high angular resolution, beyond the capabilities of the most ambitious imaging projects such as
Darwin/TPF (\cite{leger_et_al96, angel_97, beichman99}). On the contrary, we show in these papers that angularly unresolved rings
and satellites are detectable by the transit method and by the same imaging and polarimetric detection projects in preparation
for extrasolar planets. In this paper we start the study of the detectability of rings.

The detection of rings would not constitute just an astrophysical curiosity, it would have some impact on the
understanding of planetary systems and on the strategy for their detection.

- {\it Dynamics of planetary systems.}
The existence of a ring would reveal indirectly the existence of satellites around the planet. Indeed the latter
(so-called confining or {\it shepherd} satellites) confine the rings
and prevent them to dilute rapidly.
The mass of a confining satellite can be extremely small. For instance, the $3.4\times10^{-7}M_{\oplus}$
satellite Janus confines the outer edge of the Saturn A ring. In the absence of
resonances, and depending on the ring viscosity, the confinement is, for a
satellite mass $M_S$ and distance to the planet $d_S$,
proportional to $M_S^2/d_S^{4}$ (\cite{borderies_82}). In this sense, the
detection of rings can be considered as an indirect detection of extremely
small extrasolar bodies.

Satellites of giant planets in the habitable zone are also of interest because they are potential abodes of life (\cite{williams_et_al1997}).
In addition, it will be interesting to search for a correlation between the presence of rings and the planet orbit in order to understand if migration
has an impact on the formation of rings: For instance a correlation of the frequency of rings with closer
Jupiters could reveal that migration favors their formation (by a stronger disk-protoplanet mass exchange), while an anti-correlation could reveal
that migration inhibits ring or satellite formation. These are just suggestions to be investigated by detailed formation models.

- {\it Strategy for planet detection.}
The radius $r_p$ of planets detected in the thermal infrared by the future Darwin/TPF space mission is expected to be inferred from the thermal
flux $F_{p,IR}$ through the relation $F_{p,IR}\propto r_p^2$ (\cite{schneider01}, 2002). However, the latter relation holds only for spherical bodies,
so that the presence of a ring (non-spherical by definition) would give a wrong value for $r_p$. In
particular, one cannot exclude that small (e.g. Earth-like) planets have giant rings (Sect. \ref{ringed_earth}).
In addition, as will be explained in the discussion (Sect. \ref{discussion}), a ring may hide
the planet in some configurations during some parts of its orbital revolution (\cite{desmarais_et_al2002a}, 2002b).
We discuss in Sect. \ref{large and thick ring} the operational consequences of this effect: In reflected light,
if there is no planet detection at a given period of observation, this non-detection does not constitute a proof that there is no planet.

It is expected that the next generation of space or ground-based high-contrast imaging instruments (with adaptive optics,
coronagraphy, pupil apodization) will be able to deliver the first image of an unresolved extrasolar planet.
Two options are currently under consideration for TPF: thermal infrared and visible light. The emitted thermal infrared light
from the planet should show no phase effect assuming the planet is in thermal equilibrium. But the reflected visible light will vary
with phase angle, as should be shown by a broad-band photometric follow-up of the planet during its orbital motion.

Therefore we argue that it is of interest to study how the presence of a ring around a planet would influence its
magnitude as a function of its orbital position. The aim of this paper is to show that the reflected light curve of a ringed planet
is significantly different from that of a ringless planet, thus revealing the presence of a ring. The specific signature of a ring was discussed
by Schneider (2001).
Here a more quantitative model is discussed, in which the basic optical properties of the
planet and the ring are taken into account, together with all geometrical parameters describing the ringed planet.

In Sect. \ref{section_ring}, we first briefly review the known rings in the Solar System. In Sect. \ref{section_model}, we
describe our numerical model and list our different assumptions. Sect. \ref{discussion} discusses first the case of the photometry
of a lambertian ringless planet, so that we can always refer to this basic case during the discussion of our ringed
planet simulations. We finally show selected sets of light curves of a ringed planet for different geometrical and optical
parameters that demonstrate the specific photometric signature of a ring around an extrasolar planet. Following papers
will discuss the detection of satellites and rings by direct imaging, transits, astrometry and polarimetry.

\section{Rings in the Solar System}
\label{section_ring}

All giant planets of the Solar System have rings, ringlets or arclets. Those of
Saturn are brighter than the planet at visible wavelengths, while those of Jupiter, Uranus and Neptune are much fainter.

At methane absorption wavelengths, the albedo of these planets is lower than in the visible: Rings appear brighter, as for Saturn,
where the planet reflectivity becomes one order of magnitude lower than the ring's almost unchanged reflectivity
(\cite{clark_cord1980,allen1983,prinn_et_al1984}, \cite{poulet2002}). For Uranus, the rings of geometric albedo
$\approx0.05$ (\cite{cuzzi1985,karkoschka1997}) appear with a brightness comparable to the planet in the $CH_4$ band (\cite{allen1983,
roddier_et_al1998, karkoschka1998,lellouch_et_al2002}). In contrast, if emitted rather than reflected light is considered,
the atmosphere of Saturn at $4.8\mu m$ becomes transparent, allowing internal heat to escape, thus making the planet much brighter than
the ring (\cite{allen1983}).

Rings around Earth-like planets must also be considered. Indeed several authors suggest that our planet might have been
surrounded by a ring during $10^5$ to several $10^6$ years
(Myr). The assumption of a temporary orbiting debris ring following a large object impact on Earth
is made to explain at least two Earth glaciations, the ring shadowing and cooling the winter hemisphere.
This theory is proposed for the cooling which starts just after the Eocene/Oligocene boundary 35 Myr ago (\cite{okeefe1980a},
\cite{okeefe1980b}). Although the date of the cooling and the age of Eocene/Oligocene identified impact craters
do not  correspond stratigraphically (\cite{claeys2003}), a ring-induced cooling mechanism remains possible
(\cite{fawcett2002}).
The excess of $^3$He in the sediment of that period is interpreted as an abnormally high cometary bombardment rate
lasting 2.5 Myr (\cite{farley_et_al1998}). This may have created temporary rings.

The Neoproterozoic global glaciation about 590 Myr ago is also supposed to have been induced by a debris ring
(\cite{crowell1983}, \cite{fawcett2002}); this event may be connected to an impact structure in Australia
called the Acraman crater (\cite{williams1994}).

At the very beginning of the Solar System, around 4.4 Gyr ago, the Earth was impacted by a Mars-size object.
Current simulations show that the resulting debris disk condensed into the Moon core in less than one year
(\cite{takeda_ida2001}, \cite{morbidelli2003} review). But we do not know if an uncondensed residual debris disk of
1/1000th of the Moon mass, although corresponding to the actual mass of the Saturn ring, might last much longer, outside or inside the
Earth Roche lobe, than the Moon core condensation time.

We consider that a ring might be a common feature of an extrasolar planet, at least for giant planets, since for the Earth
a ring might have been present for only about $1/1000$th of the time since its formation.

\section{The ringed planet model}
\label{section_model} 
\begin{figure}[t]
   \centering
   \includegraphics[width=8.75cm]{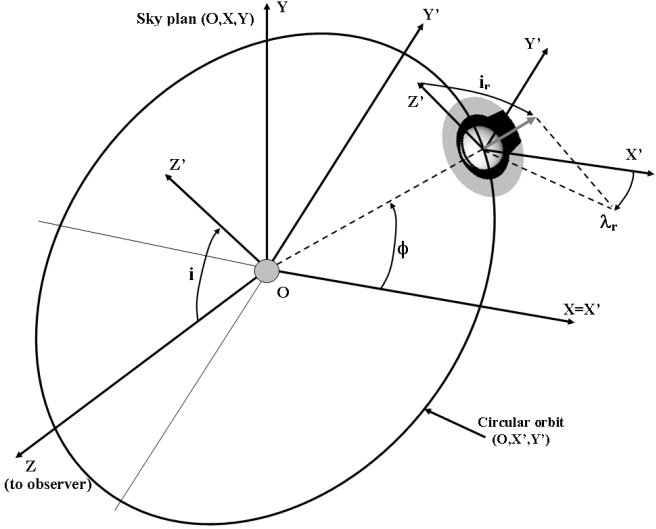}
   \caption{Reference orthogonal frames and angles definition. Angles are positive in the counterclockwise direction.
   The direction $(O,Z)$ is towards the observer. The angle $i$ defines the inclination of the real orbit in $(O,X',Y')$ with respect to the
   projected orbit on the sky $(O,X,Y)$. Here we have $i<0$. The angle $\phi$ represents the orbital phase of the planet along its circular orbit.
   The obliquity $i_r$  and longitude $\lambda_r$ define the orientation of the ring normal. Here we have $\lambda_r<0$.}
   \label{exoring-fig1}
\end{figure}

Nine photometric and geometric parameters are used to define the ringed planet (Table \ref{param} and Fig.
\ref{exoring-fig1}) and several assumptions are made to derive theoretical light curves of the ringed planet along its orbit.

   \begin{table}
      \caption[]{Geometric and optical parameters of the model.}
         \label{param}
     $$
         \begin{array}{p{0.6\linewidth}l}
            \hline
            \noalign{\smallskip}
            parameter      &  $notation$ \\
            \noalign{\smallskip}
            \hline
            \noalign{\smallskip}
            Orbit inclination wrt$^{\mathrm{a}}$ sky & i \\
            Planet: \\
            \ \ Radius & r_p     \\
            \ \ Bond albedo & A_p     \\
            Ring:\\
            \ \ Inner radius & r_i     \\
            \ \ Outer radius & r_o     \\
            \ \ Single scattering albedo (Bond)&  \varpi_0   \\
            \ \ Normal optical thickness & \tau \\
            \ \ Obliquity & i_r \\
            \ \ Ring inclination longitude & \lambda_r \\
            \noalign{\smallskip}
            \hline
         \end{array}
     $$
     $^{\mathrm{a}}$: with respect to.
   \end{table}

A circular orbit is considered for simplicity. An elliptical orbit adds a modulation proportional to the major/minor axis ratio squared, if one
considers the reflected flux versus the orbital phase angle $\phi$. Astrometric measurements of the discovered extrasolar planet
will give the orbital parameters, making possible the derivation of the measured magnitude as a function of the orbital phase $\phi$,
rather than the position angle measured on the sky. This will simplify the comparison of theoretical photometric curves of a ringless
or ringed planet with the data.

The illuminating parent star is assumed
to be at infinity and unresolved from the planet. The planet thus receives a flux
density of parallel radiation. This flux is taken to be unity. We define the illumination angle $\alpha_1$ as the angle seen from
a point (on the ring or the planet) between the star and the surface normal at that point. We define the emission angle $\alpha_2$ as
the angle seen from a point (on the ring or on the planet) between the observer and surface normal at that point.
Following several authors, we note the quantities $\cos(\alpha_1)=\mu_0$ and $\cos(\alpha_2)=\mu$. Thus
$\mu\mu_0>0$ means that the ring shows its illuminated side, while $\mu\mu_0<0$ means it shows its dark side.

To compute the light curve of the ringed planet as a function of the orbital phase angle, we build radiance maps of the
object (\cite{lester_et_al1979}, \cite{fairbairn2002}), since the detector irradiance from which the photometry is derived is
proportional to the radiance of the object. We assume that the planet is an isotropic (lambertian) gray diffusor. Although known
planets do not have a pure Lambert-like phase function, it is out of the scope of this paper to discuss a detailed model of the planet
atmosphere, whose brightness could be described for instance by a Minnaert law (\cite{minnaert1941}).
A Lambert sphere nevertheless is a good first approximation for a gas giant planet, as it reproduces the limb darkening
of a perfect spherical diffusor. Therefore each irradiated ($\mu_0>0$) point of the planet surface has a radiance of $f\times\mu_0$.
The bidirectional reflectance-distribution function $f$ is a constant $\gamma$ for a lambertian diffusor and is
related to the planet Bond albedo $A_p$ by $A_p=\pi \gamma$ (\cite{lester_et_al1979}). Thus an irradiated point of the planet has
a radiance
\begin{eqnarray}
\label{L_p}
L_p={A_p\mu_0\over\pi}.
\end{eqnarray}

We assume that the ring is a planar, homogeneous and anisotropic (non-lambertian) gray scattering layer. Isotropic (lambertian) scattering is observed at the
outer zone of Saturn's A ring (\cite{dones_et_al1993}). An isotropic scattering
assumption  neglects any 'opposition effect', i.e. a brightness increase near opposition due to stronger backscattering,
which is $\approx0.25$ magnitude in V when $\alpha<2^{\circ}$ in the case of Saturn's rings (\cite{franklin_cook1965}). The phase
angle $\alpha$ is the angle seen from the planet between the star and the observer.

In first approximation, the ring brightness is estimated by assuming only single scattering in the ring.
Multiple scattering would require a much more elaborate ring model (particle size distribution, spatial distribution along the ring
thickness, etc.).
Diffuse reflected and transmitted radiances from single scattering are respectively,
per unit incident flux density (\cite{chandra1960}),
\begin{eqnarray}
\label{I_R}
I_R = {\varpi_0\ |\mu_0|\over 4(|\mu|+|\mu_0|)}
\big[1-\exp\big(-\tau({1\over|\mu|}+{1\over|\mu_0|})\big)\big] \\
= {\varpi_0\ |\mu_0|\over 4(|\mu|+|\mu_0|)}
\big[1-\exp(-\tau_{path})\big]
\label{I_R2}
\end{eqnarray}
with $\tau_{path}=\tau({1\over|\mu|}+{1\over|\mu_0|})$, and
\begin{eqnarray}
\label{I_T}
I_T = {\varpi_0\ |\mu_0|\over 4(|\mu|-|\mu_0|)}
\big[\exp(-\tau/|\mu|)-\exp(-\tau/|\mu_0|)\big]
\end{eqnarray}
if $|\mu|\neq|\mu_0|$, and
\begin{eqnarray}
I_T = {\varpi_0\ \tau\over 4|\mu_0|}\exp(-\tau/|\mu_0|)\ {\rm if}\
|\mu|=|\mu_0|.
\end{eqnarray}

Mutual lighting (i.e. planet-shine on the ring, or ring-shine on the planet) is neglected here. It is negligible especially if one
of the bodies (planet or ring) has an albedo significantly lower than the other, or if both have low albedos.
But at high phase angles
(planet seen as a thin crescent) and when the ring does not show its illuminated side, our model may compute a fainter
magnitude than reality, because we do not compute the planet-shine on the observed dark side of the ring.
For Saturn, this planet-shine flux can be comparable to the light flux
transmitted by the ring (\cite{smith_et_al1982}, \cite{poulet_et_al2000}).

Mutual shadowing is computed because with rings of large optical thickness and for given illumination and emission angles, the ring
projects its shadow on the planet making the planet parts in the shadow of the ring fainter. If we consider a unit incident flux on the ring, then the transmitted
flux in the incident direction is $\exp(-\tau/|\mu_0|)$. Thus
a planet region in the shadow of the ring has a radiance
\begin{eqnarray}
\label{L_ps}
L_{ps}= L_p\exp(-\tau/|\mu_0|).
\end{eqnarray}
The planet parts also appear fainter when they are observed through the ring. Following the same reasoning
as above, these regions have a radiance
\begin{eqnarray}
\label{L_pr}
L_{pr}= L_{ps,p}\exp(-\tau/|\mu|)
\end{eqnarray}
where $L_{ps,p}=L_{ps}$ or $L_p$ depending of whether the region is in the shadow of the ring or not, respectively.
The planet also projects its shadow on the ring and this can correspond to a significant part of the
ring surface (see figures below): Ring regions in the shadow of the planet are set to zero.

The planet is spherical for simplicity. The oblateness of Saturn is $\approx0.1$ which represents a magnitude difference
of less than $\approx 0.1$ with respect to a spherical planet (for photometry of ellipsoids, see \cite{fairbairn2003}).

\section{Discussion}
\label{discussion}
\subsection{The photometry of a lambertian ringless planet}
\label{section planet_no_ring}
\begin{figure}[t]
   \centering
   \includegraphics[width=8.75cm]{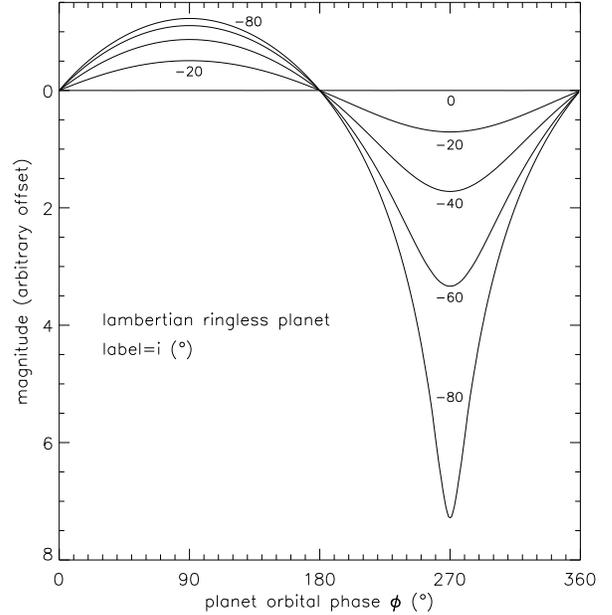}
   \caption{Magnitude variation of a ringless planet shown for different inclinations {\it i} of the
   orbital plane with respect to the sky. All curves are normalized to the flux of a ringless planet seen at half phase.
   When $i=0^{\circ}$, the phase angle is a constant equal to $90^{\circ}$ and the planet is always
   seen as a half globe of constant brightness (because the orbit is circular in our model). For positive values of {\it i}, the graph is
   the time-inverse of the one plotted and we have for a ringless planet: magnitude$(-i,\phi)$ = magnitude$(i,-\phi)$.}
\label{planet_no_ring}
\end{figure}
Fig. \ref{planet_no_ring} shows the magnitude versus the orbital phase $\phi$ for a ringless planet observed at different inclinations $i$ ranging
from 0 to $-80^{\circ}$. For $i=0^{\circ}$ (pole-on view), $\alpha$ is always equal to $90^{\circ}$ and the magnitude is a constant. For
$i=\pm90^{\circ}$ (edge-on view), all phase angles can in principle be probed except around $\alpha=0^{\circ}$ where the planet is behind the star and around
$\alpha=180^{\circ}$ where the planet transits. In our reference frame (Fig. \ref{exoring-fig1}), for all $i$ we have $\alpha=90^{\circ}$ at $\phi=0^{\circ}$.
For a given $i\neq0$,
the minimum $\alpha$ (full or near full phase)  is reached at $\phi=90^{\circ}$ and maximum $\alpha$ at $\phi=270^{\circ}$.
Therefore, with our assumption of a circular orbit, the planet maximum brightness occurs at $\phi=90^{\circ}$ and minimum at $\phi=270^{\circ}$.

It must be noted here that the lambertian planet brightness changes by a factor $>\approx100$ for systems seen at high
inclination (Fig. \ref{planet_no_ring}). These brightness extrema occur when the position of the planet projected on the sky becomes closer to the star: The planet may
disappear in the residual star halo, especially at high phase angle when it becomes a thin and faint crescent. The planet will thus
remain below the detection level for a long time but this obviously would not constitute a proof of the absence of planets
around that star. From an operational point of view, this means that a given star must be monitored over a period of time of the order
of the expected orbital period, especially if it is suspected that the system is observed at high inclination.

\subsection{The photometry of a lambertian ringed planet}
\subsubsection{ A Saturn-like planet}
\begin{figure}
   \centering
   \includegraphics[width=8.75cm]{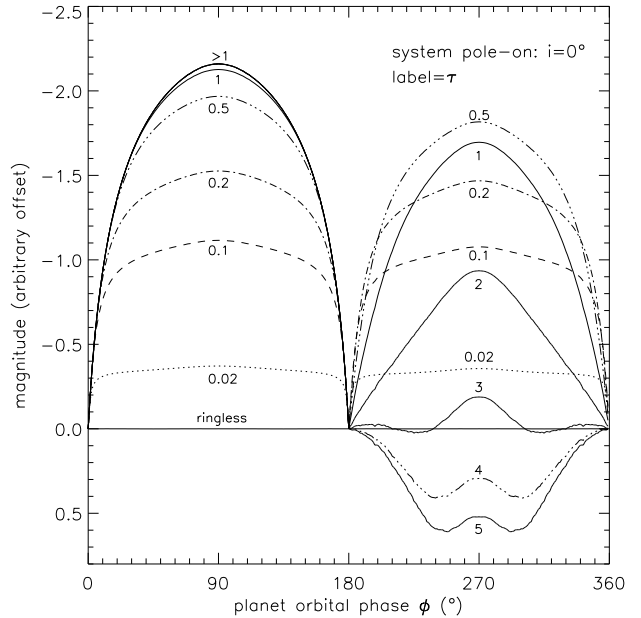}
   \caption{Pole-on ringed planet photometry. A Saturn-like planet is considered, with different optical thicknesses of the ring (Table \ref{values}).
   The planet is always seen at half phase and the ring shows its illuminated face during the first half-orbit.
   During the second-half, the system brightness results from the light transmitted through the ring, but also from the planet partially dimmed by the ring shadow.
   For a thick ring ($\tau>\approx3$),
   the transmitted light is negligible and the light curve variations are dominated by the effect of the ring shadow on the planet.
   The curve $\tau=2$ is illustrated by Fig. \ref{ap34_k100_rp30_ar70_tau200_ri45_ro69_ir26_lambdar-90_i0_step1_image_along_orbit}.
   All curves are normalized to the flux of a ringless planet seen at half phase.}
\label{system_pole_on_ir26}
\end{figure}

\begin{figure}
   \centering
   \includegraphics[width=8.75cm]{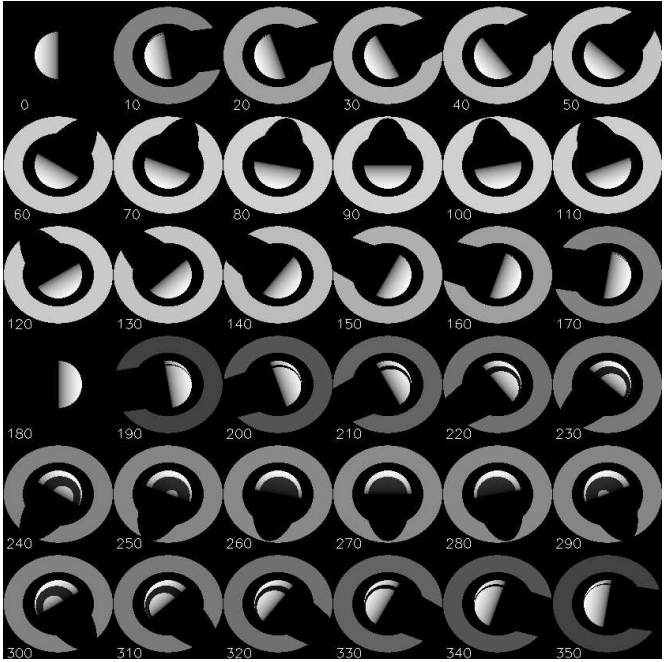}
   \caption{Pole-on planet for different orbital positions $\phi$ (labels). Since $i=0^{\circ}$, the planet is always seen at half phase,
   while the ring, with an obliquity of $i_r=26.73^{\circ}$ (Saturn-like, Table \ref{values}), shows its illuminated face during the first half-orbit.
   During the second half-orbit,
   the back-illuminated ring becomes fainter. The corresponding light curves are given in Fig. \ref{system_pole_on_ir26}.
   For figure clarity, here $\tau=2$, thicker than Saturn's ring. Image scale = radiance$^{0.33}$.}
\label{ap34_k100_rp30_ar70_tau200_ri45_ro69_ir26_lambdar-90_i0_step1_image_along_orbit}
\end{figure}

\begin{figure}
   \centering
   \includegraphics[width=8.75cm]{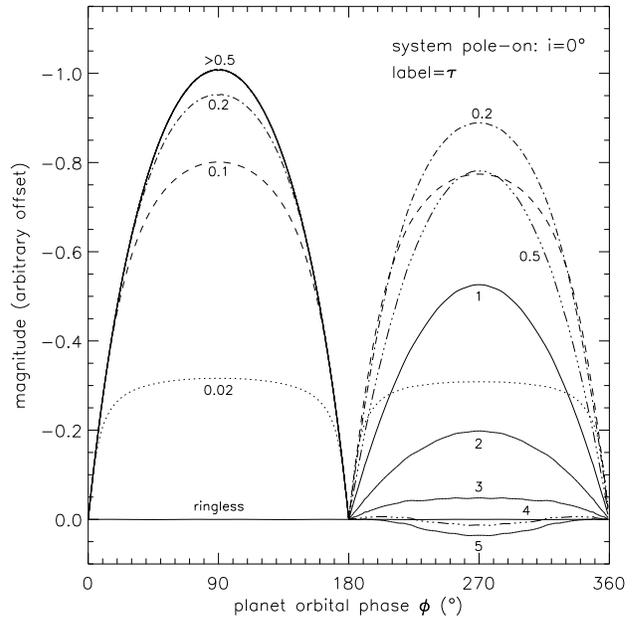}
   \caption{Pole-on ringed planet photometry. We consider here a Saturn-like planet, but with a lower obliquity of $i_r=5^{\circ}$ (Table \ref{values}).
   The planet is always seen at half phase and the ring shows its illuminated face during the first half-orbit.
   During the second half-orbit, the brightness increases again thanks to the light transmitted through the ring. But for a thick ring ($\tau>\approx3$),
   the transmitted light is negligible and the light curve becomes almost flat, dominated by the planet seen at half phase,
   the ring shadow being very small due to the low obliquity. Note also that the overall magnitude range is lower for $i_r=5^{\circ}$ than
   for $i_r=26.73^{\circ}$ because higher ring illumination angles (i.e. higher $|\mu_0|$) are reached when $i_r$ increases, for a given inclination $i$.
   All curves are normalized to the flux of a ringless planet seen at half phase.}
\label{system_pole_on_ir5}
\end{figure}

Let us first consider a Saturn-like planet (Table \ref{values}).
When the system is observed pole-on, the planet is always seen at half phase ($\alpha=90^{\circ}$), while the ring alternately shows
its illuminated and dark side.
The light curves of Fig. \ref{system_pole_on_ir26} thus show slope changes occurring at the equinoxes, due to the two observation regimes of the ring, seen either
reflecting or transmitting the light.  For Saturn, with typically $\tau=1$, the curve is almost symmetrical: Reflected and transmitted radiances are of
the same order of magnitude. But for a thicker ring, the curves in Fig. \ref{system_pole_on_ir26} become
highly non-symmetrical as illustrated in Fig. \ref{ap34_k100_rp30_ar70_tau200_ri45_ro69_ir26_lambdar-90_i0_step1_image_along_orbit}: The reflected radiance is much larger
than the transmitted one. The ring shadow projected on the planet is responsible for brightness changes too,
as shown especially for high optical thickness. Moreover, with a low-obliquity ring, the transmitted light {\it a fortiori} remains low and the ring
shadow on the planet becomes small, making the system brightness almost constant for the second half of the orbit (Fig. \ref{system_pole_on_ir5}).
Note that the reflected light is almost independent of $\tau$, because the exponential term in $I_R$ (Eq. \ref{I_R2}) becomes small with respect
to 1 when $\tau_{path}>1$. The light already intercepts many particles, so adding or removing more particles does not change the ring radiance
significantly (\cite{dones_et_al1993}).

Fig. \ref{system_pole_on_ir26} and \ref{system_pole_on_ir5} thus show that, when a ring is present, the light
curve presents four extrema, instead of two for a ringless planet (Fig. \ref{planet_no_ring}).
This light curve dichotomy was already identified as a ring photometric signature and qualitatively
discussed by Schneider (in \cite{desmarais_et_al2002a}).
\begin{table}
      \caption[]{Parameters for different ringed planets. Definitions are given in Table \ref{param} and Fig. \ref{exoring-fig1}.}
         \label{values}
     $$
         \begin{array}{p{0.4\linewidth}lll}
            \hline
            \noalign{\smallskip}
            parameter       &  $Saturn-like$  & $Large and$ & $Ringed $\\
                            &                 & $thick ring$ & $Earth-like$\\
            \noalign{\smallskip}
            \hline
            \noalign{\smallskip}
            $A_p$          & 0.34            &  0.34        & 0.30\\
            $A_p\ (CH_4)$  & 0.05 $\ to\ $ 0.15  &  -       & - \\
            $r_i$          & 1.5\ r_p        &  1.1\ r_p    & 1.3\ r_p\\
            $r_o$          & 2.3\ r_p        &  2.7\ r_p    & 10\ r_p \\
            $\varpi_0$     &  0.7  & 0.7 $\ or\ $ 0.05     & 0.7 $\ or\ $ 0.05\\
            $\tau$         & 1               &  1 $\ to\ $ 5& 0.02 $\ to\ $ 0.5\\
            $i_r$          & 26.73^{\circ}   & 40^{\circ}   & 23.43^{\circ}\\
            $\lambda_r$      & -90^{\circ}     & -40^{\circ}  & 30^{\circ}\\
            \noalign{\smallskip}
            \hline
         \end{array}
     $$
   \end{table}



\begin{figure}
   \centering
   \includegraphics[width=8.75cm]{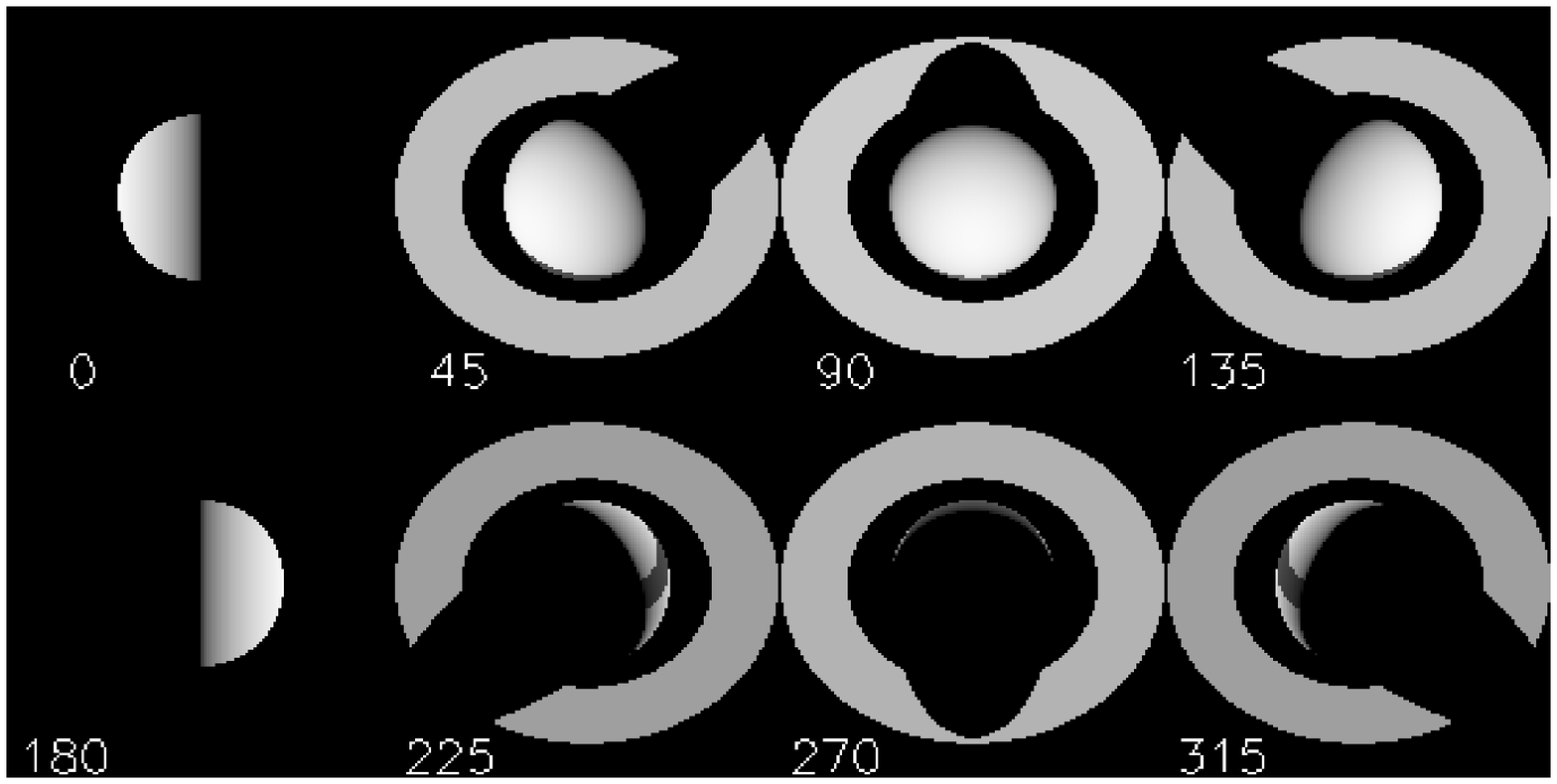}
   \includegraphics[width=8.75cm]{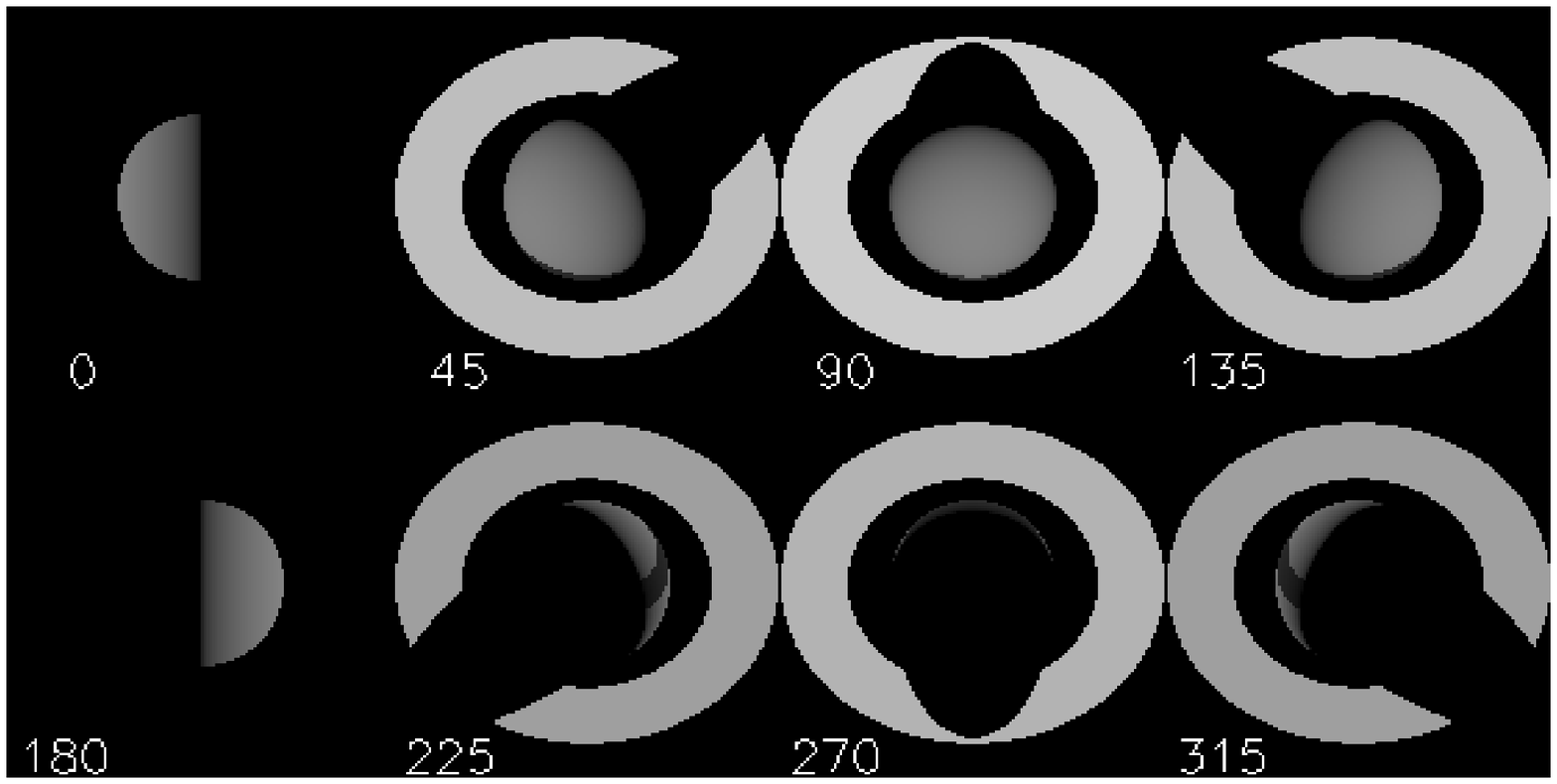}
   \caption{Saturn-like planet for different orbital positions $\phi$ (labels). The picture above represents the planet in the continuum ($A_p=0.34$)
   and below in a $CH_4$ absorption band ($A_p=0.05$). In the latter case, the figure shows that the main contribution
   to the total light comes from the ring. The corresponding light curves are given in Fig.
   \ref{continuum_ap34_et_CH4_k100_rp30_ar70_tau100_ri45_ro69_ir26_lambdar-90_i-60_step1_flux}.
   The two sets have the same brightness scale. Inclination $i=-60^{\circ}$. Image scale = radiance$^{0.33}$. }
\label{ap34+5_k100_rp30_ar70_tau100_ri45_ro69_ir26_lambdar-90_i-60_step10_image_along_orbit}
\end{figure}
\begin{figure}
   \centering
   \includegraphics[width=8.75cm]{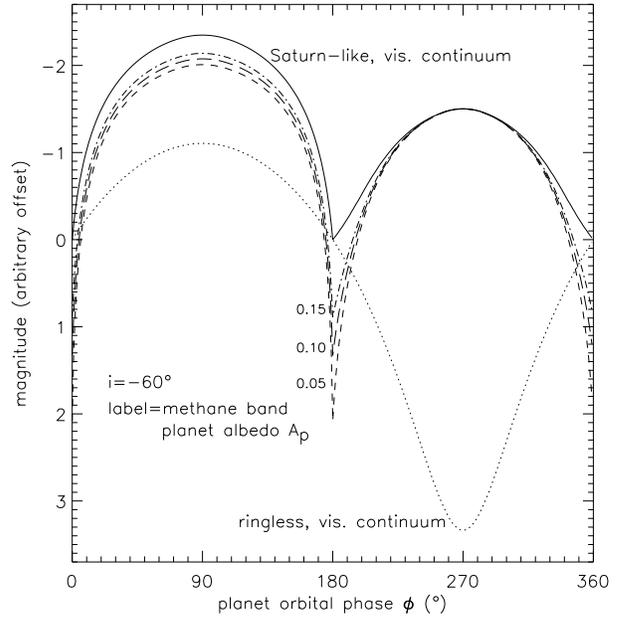}
   \caption{Saturn-like planet photometry for two different wavelengths: $CH_4$ absorption band and continuum (Table \ref{values} and
   Fig. \ref{ap34+5_k100_rp30_ar70_tau100_ri45_ro69_ir26_lambdar-90_i-60_step10_image_along_orbit}).
   Note the different variation amplitude in the $CH_4$ band and the continuum. Note also the strong slope change at the equinoxes,
   occurring here at $\phi=0^{\circ}$ and $180^{\circ}$, when the ring comes to be seen illuminated or back-illuminated, respectively.
   All curves are normalized to the flux of a ringless planet seen at half phase.}
   \label{continuum_ap34_et_CH4_k100_rp30_ar70_tau100_ri45_ro69_ir26_lambdar-90_i-60_step1_flux}
\end{figure}


\begin{figure}
   \centering
   \includegraphics[width=8.75cm]{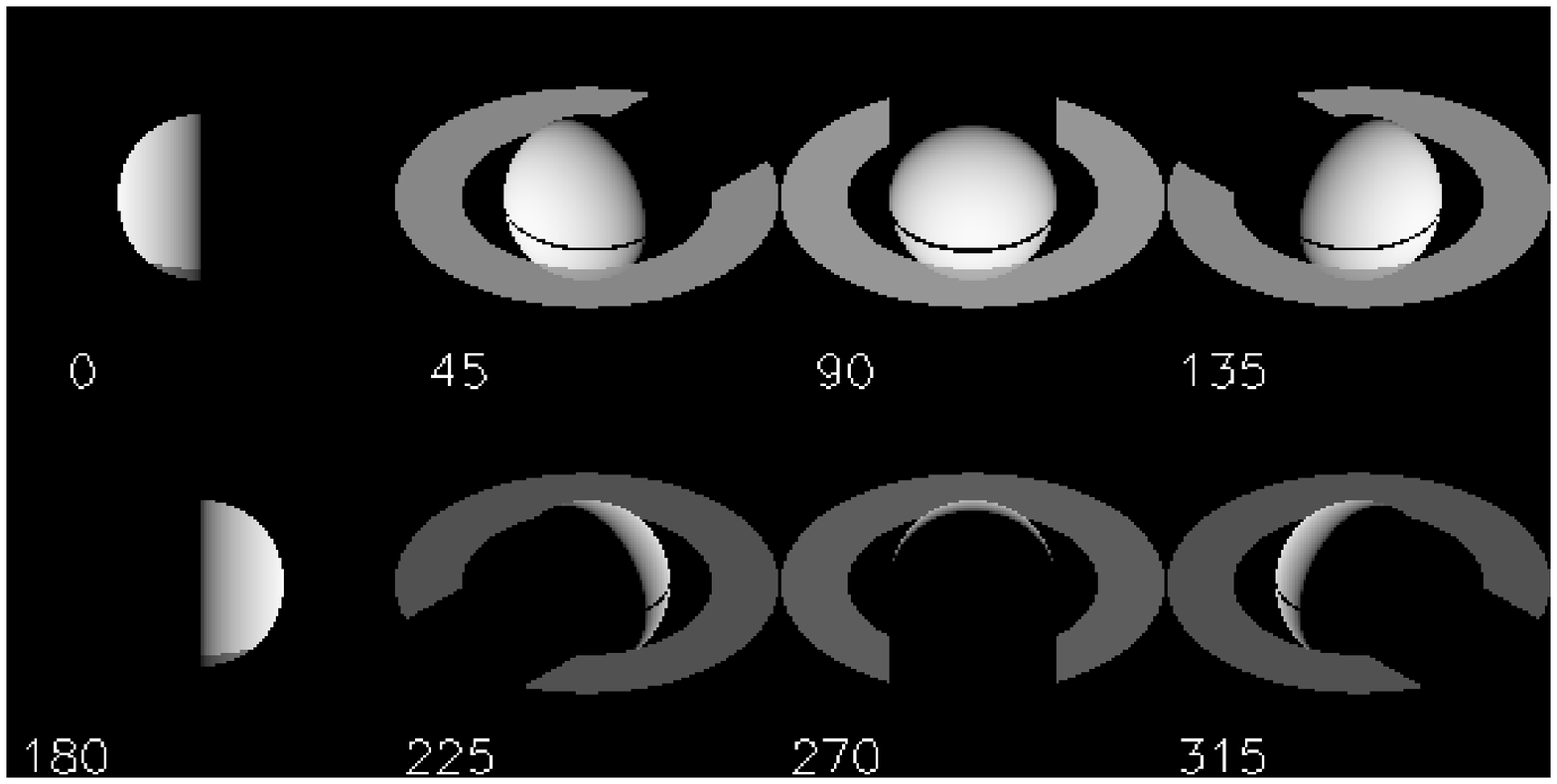}
   \includegraphics[width=8.75cm]{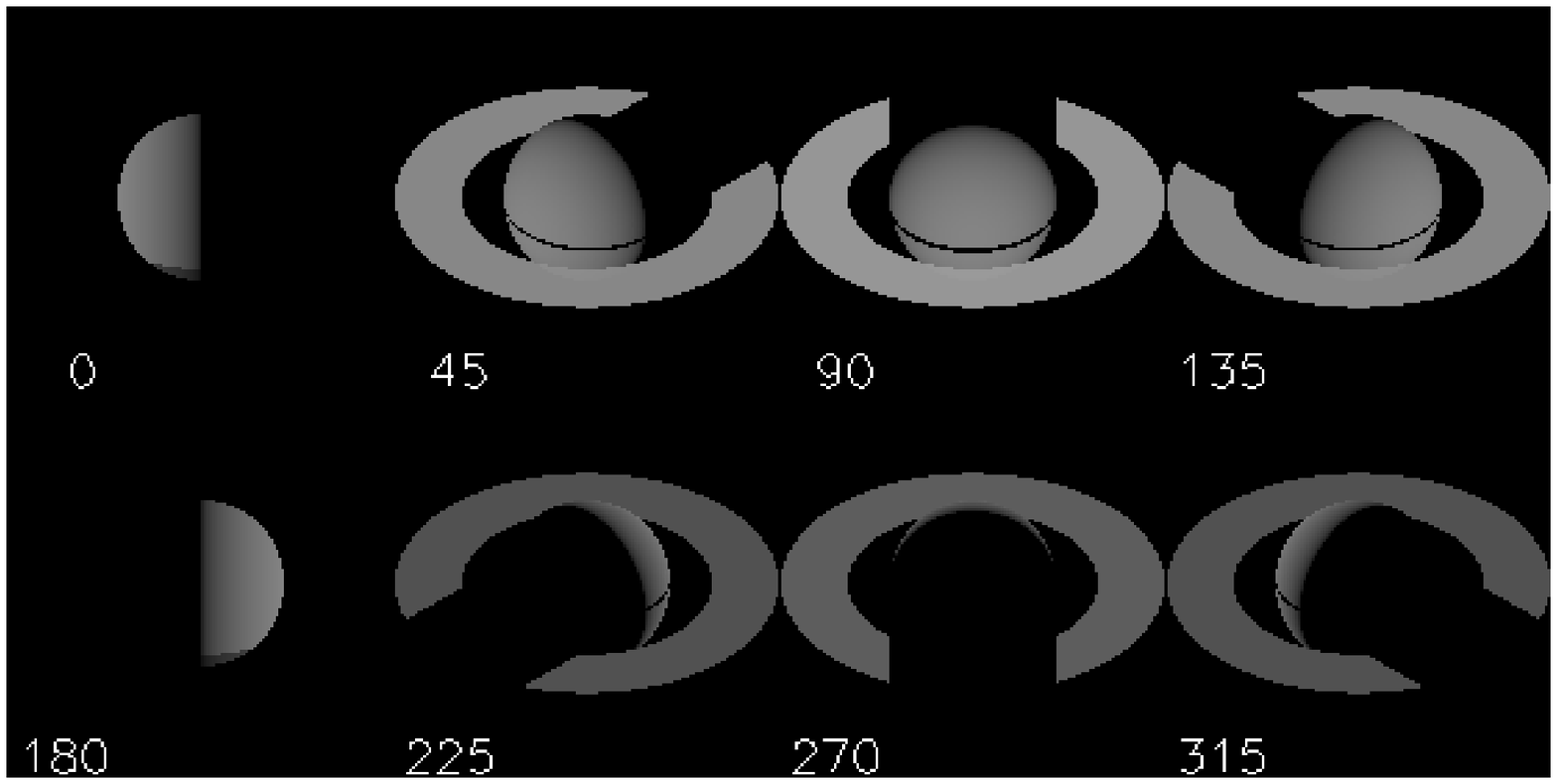}
   \caption{Ringed planet for different orbital positions $\phi$ (labels). Obliquity is $5^{\circ}$ instead of $26.73^{\circ}$ as for Saturn (Table \ref{values}).
   The picture above represent the planet in the continuum ($A_p=0.34$) and below in a $CH_4$ absorption band ($A_p=0.05$).
   The corresponding light curves are given in Fig.
   \ref{continuum_ap34_et_CH4_k100_rp30_ar70_tau100_ri45_ro69_ir5_lambdar-90_i-60_step1_flux}.
   The two sets have the same brightness scale. Inclination $i=-60^{\circ}$. Image scale = radiance$^{0.33}$. }
\label{ap34+5_k100_rp30_ar70_tau100_ri45_ro69_ir5_lambdar-90_i-60_step10_image_along_orbit}
\end{figure}
\begin{figure}
   \centering
   \includegraphics[width=8.75cm]{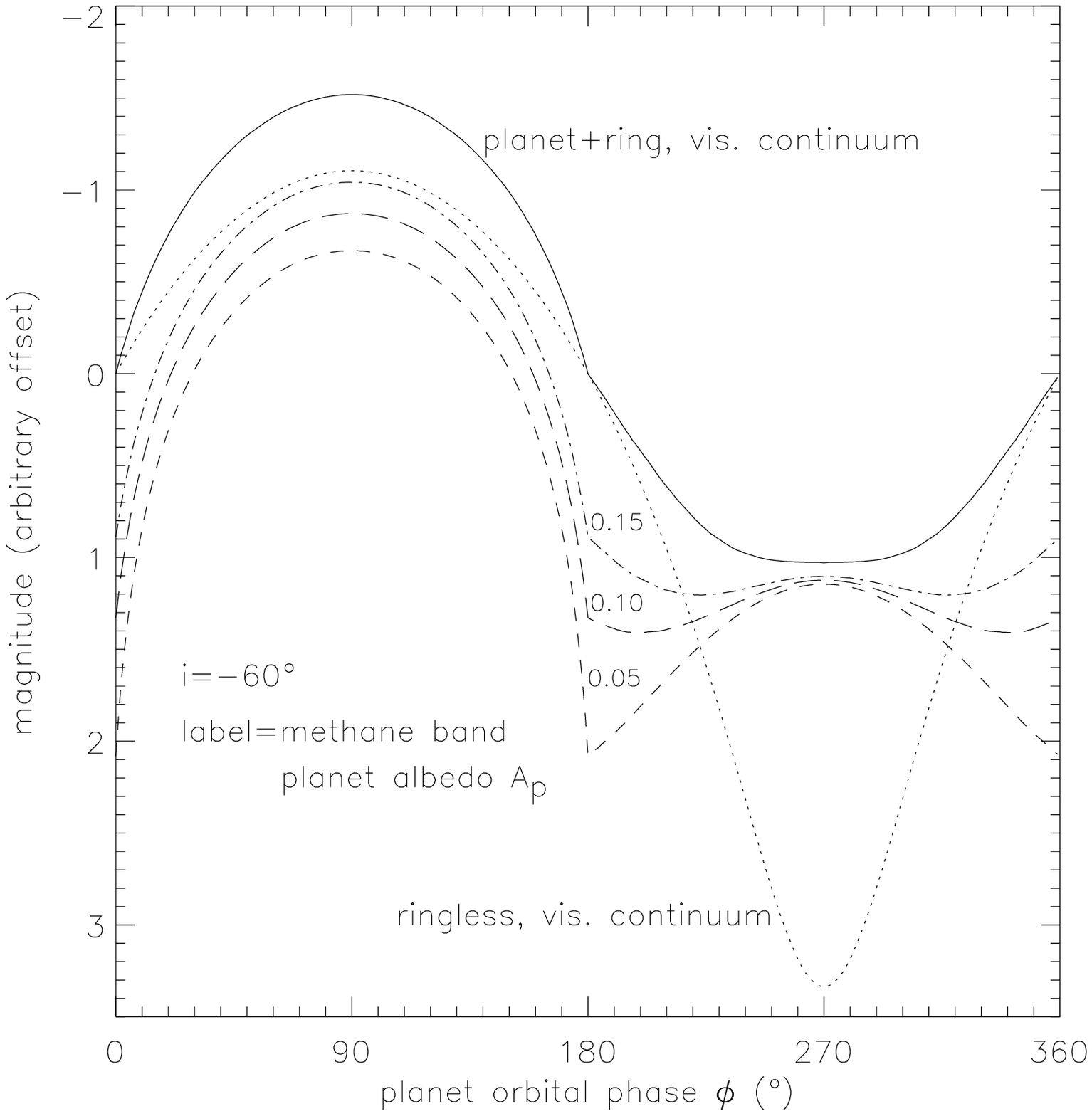}
   \caption{Ringed planet photometry for two different wavelengths: $CH_4$ absorption band and continuum. We consider here a Saturn-like planet, but
   with a lower obliquity of $i_r=5^{\circ}$ (Table \ref{values} and Fig. \ref{ap34+5_k100_rp30_ar70_tau100_ri45_ro69_ir5_lambdar-90_i-60_step10_image_along_orbit}).
   Light curves are significantly different, especially during the second half-orbit: The ring and planet vary in opposite directions and by the same order of
   magnitude, making the light curve almost flat. All curves are normalized to the flux of a ringless planet seen at half phase.}
   \label{continuum_ap34_et_CH4_k100_rp30_ar70_tau100_ri45_ro69_ir5_lambdar-90_i-60_step1_flux}
\end{figure}

The planet and the ring may have different chemical compositions, and dual-band photometry (or spectroscopy
if the object is bright enough) greatly helps to detect the ring by observing for instance in the (visible) continuum and in
a methane absorption band, where a Saturn-like planet becomes much fainter than the ring (Fig.
\ref{ap34+5_k100_rp30_ar70_tau100_ri45_ro69_ir26_lambdar-90_i-60_step10_image_along_orbit}). The corresponding photometry
(Fig. \ref{continuum_ap34_et_CH4_k100_rp30_ar70_tau100_ri45_ro69_ir26_lambdar-90_i-60_step1_flux}) shows that in the $CH_4$ band the
magnitude variation can be twice the variation in the (visible) continuum.
Note that, with $\lambda_r=-90^{\circ}$ during the second half-orbit when the phase angle
is above $90^{\circ}$, the ring brightness and planet brightness vary in opposite directions. These variations can partially compensate each other, making the light
curve almost flat, if we consider for instance a higher ring optical thickness or a lower obliquity (Fig.
\ref{ap34+5_k100_rp30_ar70_tau100_ri45_ro69_ir5_lambdar-90_i-60_step10_image_along_orbit} and
\ref{continuum_ap34_et_CH4_k100_rp30_ar70_tau100_ri45_ro69_ir5_lambdar-90_i-60_step1_flux}).

Therefore, the light curves at different wavelengths can be significantly different. In the methane band, stronger slope changes occur at the equinoxes
(here at $\phi=0$ and $180^{\circ}$, since $\lambda_r=-90^{\circ}$), here again due to the two observation regimes of the ring, either seen
reflecting or transmitting the light. Note that when the ring disappears at the equinoxes,
the object spectrum can be the spectrum of the planet only (Fig.\ref{ap34+5_k100_rp30_ar70_tau100_ri45_ro69_ir26_lambdar-90_i-60_step10_image_along_orbit}),
but it can also be composite if a part of the planet is seen through the unilluminated ring
(Fig.\ref{ap34+5_k100_rp30_ar70_tau100_ri45_ro69_ir5_lambdar-90_i-60_step10_image_along_orbit}).

\begin{figure*}
   \centering
   \includegraphics[width=8.75cm]{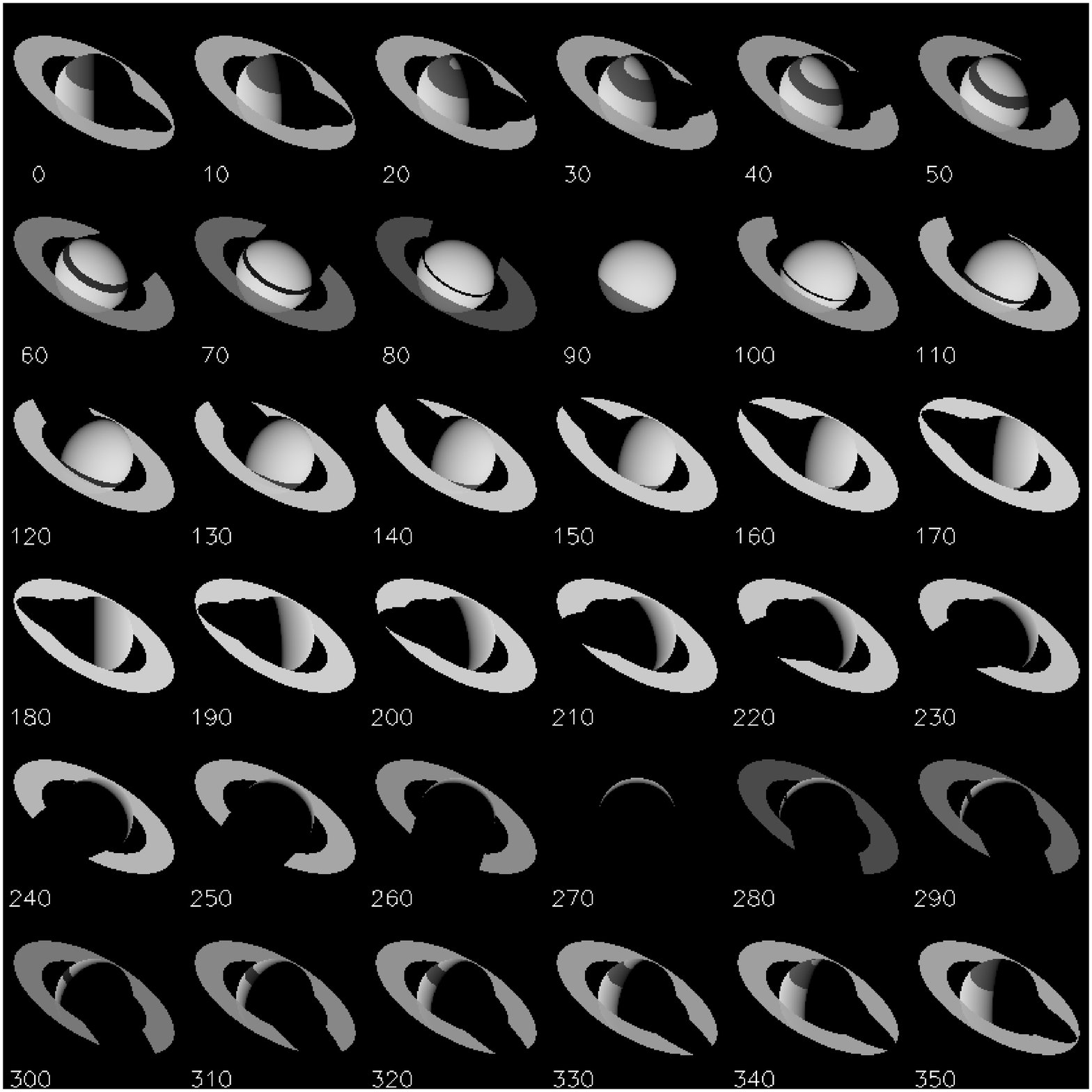}
   \includegraphics[width=8.75cm]{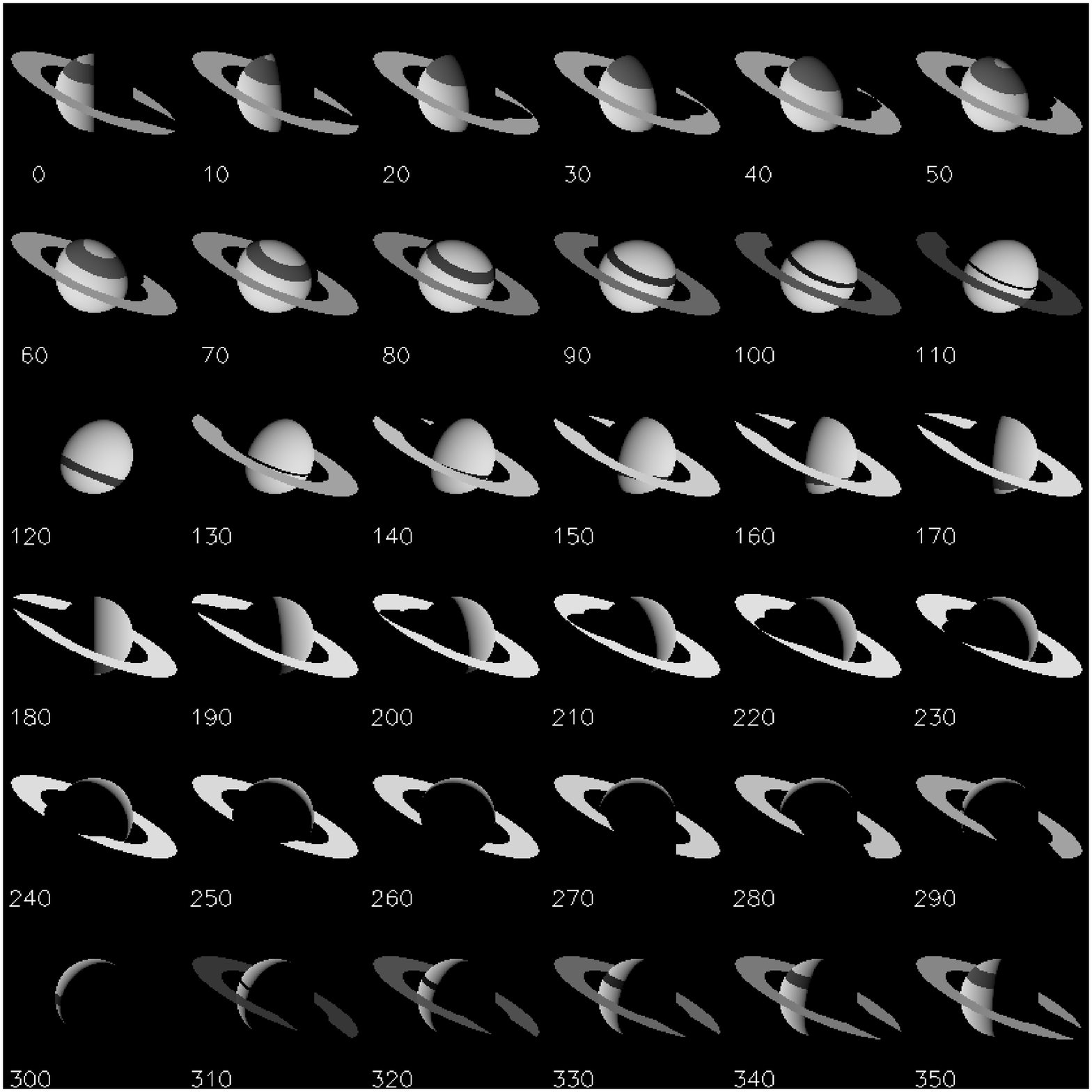}
   \includegraphics[width=8.75cm]{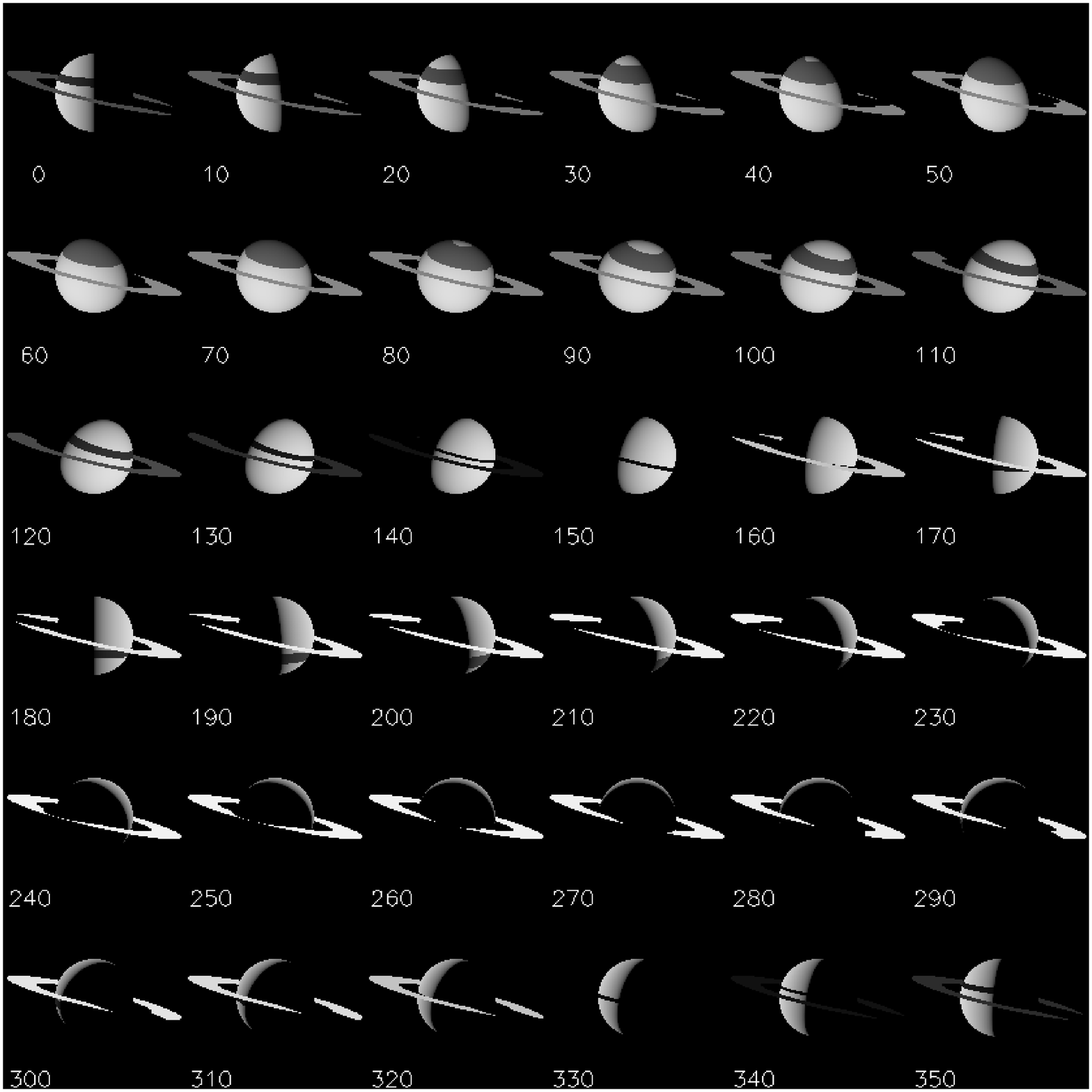}
   \includegraphics[width=8.75cm]{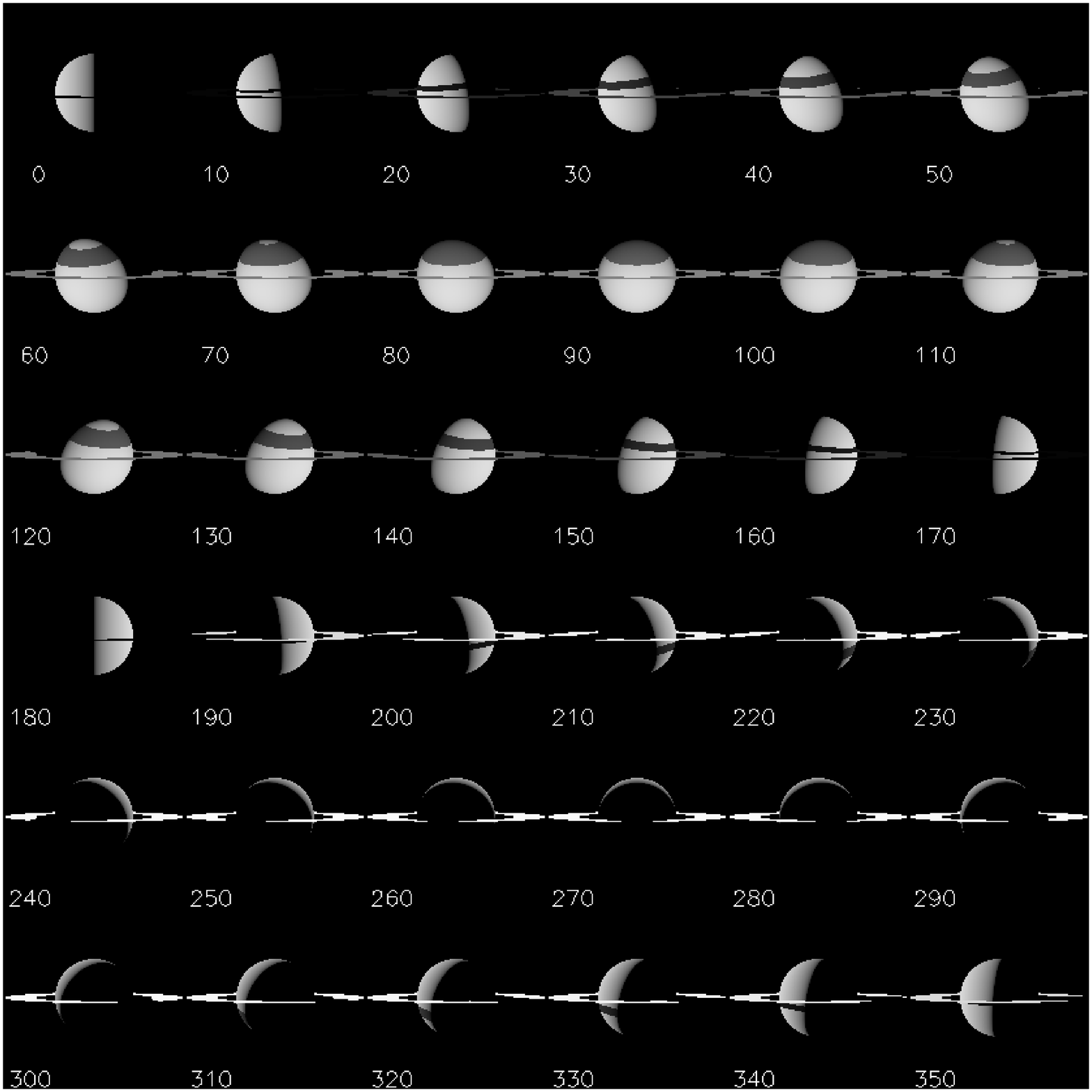}
   \caption{A Saturn-like planet for different orbital positions $\phi$ (labels). The planet (Table \ref{values}) is represented here with
   $\lambda_r=0^{\circ}$ (above left), $\lambda_r=30^{\circ}$ (above right),
   $\lambda_r=60^{\circ}$ (below left) and $\lambda_r=90^{\circ}$ (below right). This figure shows that respective ring and planet brightness extrema do
   not always occur simultaneously:
   For instance, when $\lambda_r=30^{\circ}$ (above right), a ring minimum occurs at $\phi=300^{\circ}$, while the planet minimum occurs at $\phi=270^{\circ}$.
   This difference induces a shift of the system light curve as illustrated in Fig. \ref{versus_lambdar}.
   The four sets have the same brightness scale. Inclination $i=-60^{\circ}$. Image scale = radiance$^{0.33}$.}
   \label{ap34_k100_rp30_ar70_tau100_ri45_ro69_ir26_lambdar30_i-60_step1_image_along_orbit}
\end{figure*}
\begin{figure}
   \centering
   \includegraphics[width=8.75cm]{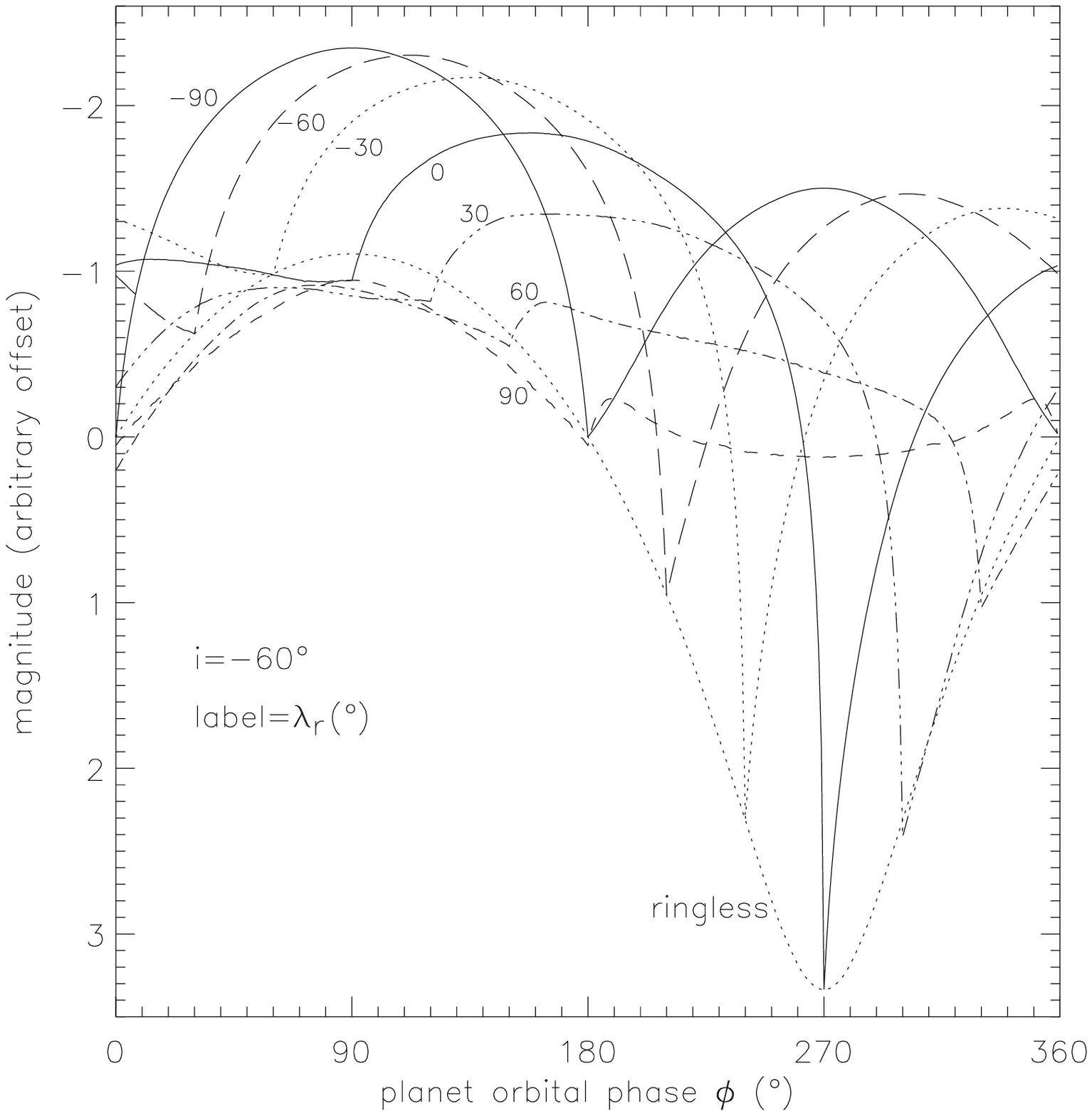}
   \caption{Saturn-like planet photometry (Table \ref{values}). Light curves for different values of $\lambda_r$, showing the shift of brightness extrema with respect to a planet
   without ring. The curve for a ringless planet is plotted for comparison. All curves are normalized to the flux of a ringless planet seen at half phase.
   Curves are plotted for $-90^{\circ}\leq\lambda_r\leq90^{\circ}$ and illustrated in
   Fig. \ref{ap34_k100_rp30_ar70_tau100_ri45_ro69_ir26_lambdar30_i-60_step1_image_along_orbit}. See also
   Fig. \ref{ap34+5_k100_rp30_ar70_tau100_ri45_ro69_ir26_lambdar-90_i-60_step10_image_along_orbit} where $\lambda_r=-90^{\circ}$.
   Note that we have the relation: magnitude$_{\lambda_r}(\phi)=$
   magnitude$_{180^{\circ}-{\lambda_r}}(180^{\circ}-\phi)$, all other parameters being unchanged.}
   \label{versus_lambdar}
\end{figure}

Let us now consider the Saturn-like planet shown in Fig. \ref{ap34_k100_rp30_ar70_tau100_ri45_ro69_ir26_lambdar30_i-60_step1_image_along_orbit}.
As soon as $\lambda_r$ is different from $\pm90^{\circ}$, the ring brightness extrema will not occur simultaneously with the brightness extrema
of the planet, i.e. at $\phi=\pm90^{\circ}$. Consequently, as Fig. \ref{versus_lambdar} shows, the brightness extrema of the ringed planet are shifted
in $\phi$ with respect to the extrema occurring at $\phi=\pm90^{\circ}$ in Fig. \ref{planet_no_ring} for a ringless planet.
We consider this {\it$\phi$-shift} of the light curve a photometric signature of a ring.
This signature is unambiguous for a ringed planet on a circular orbit. A $\phi$-shift can be observed if the orbit is elliptical, whether the planet has a ring
or not, when brightness variations are induced by distance changes from the planet to its star.
But we assume that the astrometry of the discovered planet will make possible the correction of the measured reflected light photometry for the effect of orbit ellipticity.

\subsubsection{ A planet with a larger and thicker ring }
\label{large and thick ring}
\begin{figure}
   \centering
   \includegraphics[width=8.75cm]{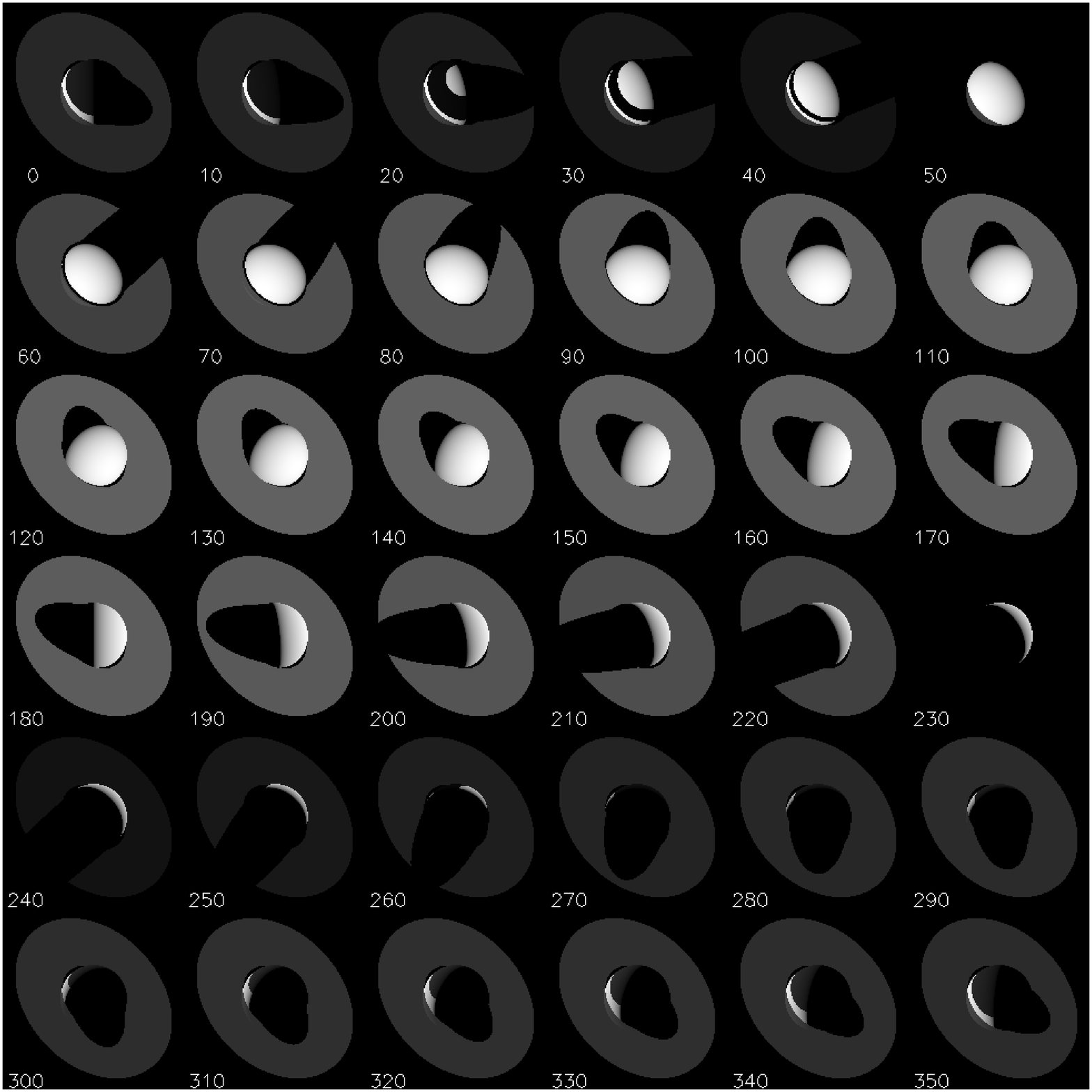}
   \caption{Ringed planet for different orbital positions $\phi$ (labels). Here the ring inner edge is close to
   the planet and consequently this large hides the planet around $\phi=270^{\circ}$ (Table \ref{values}, here $\tau=3$ and $\varpi_0=0.05$). Light curves showing the
   behavior of this system are given in Fig. \ref{anneau_saturnien_et_grand} and \ref{anneau_saturnien_et_grand_et_sombre}.
    Inclination $i=-60^{\circ}$. Image scale = radiance$^{0.33}$.}
   \label{ap34_k100_rp30_ar5_tau400_ri33_ro81_ir40_lambdar-40_i-60_step1_image_along_orbit}
\end{figure}

\begin{figure}
   \centering
   \includegraphics[width=8.75cm]{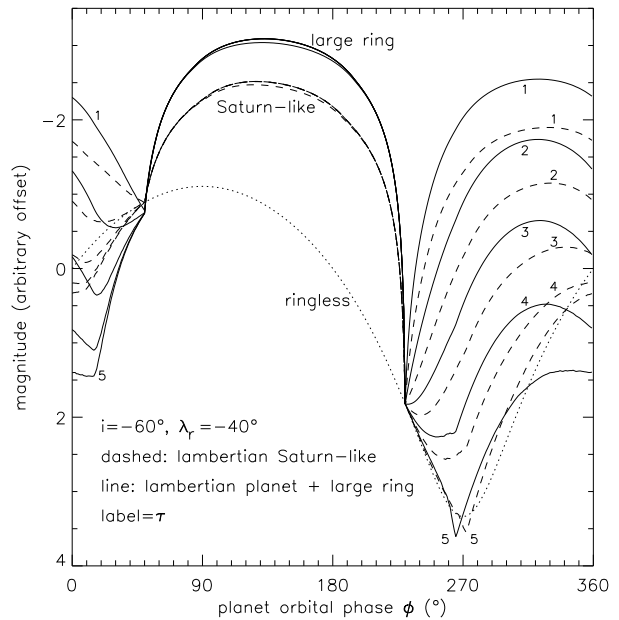}
   \caption{Ringed planet photometry versus $\phi$ for a large ring of albedo $\varpi_0=0.7$ and different $\tau$ values (Table \ref{values} and
   Fig. \ref{ap34_k100_rp30_ar5_tau400_ri33_ro81_ir40_lambdar-40_i-60_step1_image_along_orbit}).
   The ring hides the planet with a maximum around $\phi\approx265^{\circ}$.
   Curves for a Saturn-like planet with $i_r=40^{\circ}$ and $1\le\tau\le5$ are shown for comparison.
   All curves are normalized to the flux of a ringless planet seen at half phase.}
   \label{anneau_saturnien_et_grand}
\end{figure}

We now consider a planet with a ring larger than that of Saturn (Table \ref{values}).
With the chosen $i_r$ and $\lambda_r$ angles, the sequence in Fig. \ref{ap34_k100_rp30_ar5_tau400_ri33_ro81_ir40_lambdar-40_i-60_step1_image_along_orbit}
starts with the back-illuminated ring projecting its shadow over the planet polar region. For a given obliquity, the smaller the gap between
the inner edge of the ring and the planet equator, the lower the latitude of the shadow on the planet.
Also, the larger the ring outer diameter, the longer the polar region remains in the
ring shadow. The ring thus hides the planet: The light curves (Fig. \ref{anneau_saturnien_et_grand}) show that for a thick ring with $\tau>\approx3$, the light
from the planet blocked by the ring becomes stronger than the light transmitted by the ring, making the system fainter than a ringless planet.
A thick and large ring can almost completely hide the planet, as shown here around
$\phi=270^{\circ}$ in Fig.
\ref{ap34_k100_rp30_ar5_tau400_ri33_ro81_ir40_lambdar-40_i-60_step1_image_along_orbit}. Note that when the planet is hidden, the object spectrum
is dominated by the ring spectrum.

On the other hand, when $\approx50<\phi<\approx230^{\circ}$, the illuminated side
of the ring is visible and induces a significant increase of brightness of $\approx2$ magnitudes of the system when the ring has a high albedo.
At both equinoxes, here again, we observe a strong slope change of the light curves.

\begin{figure}
   \centering
   \includegraphics[width=8.75cm]{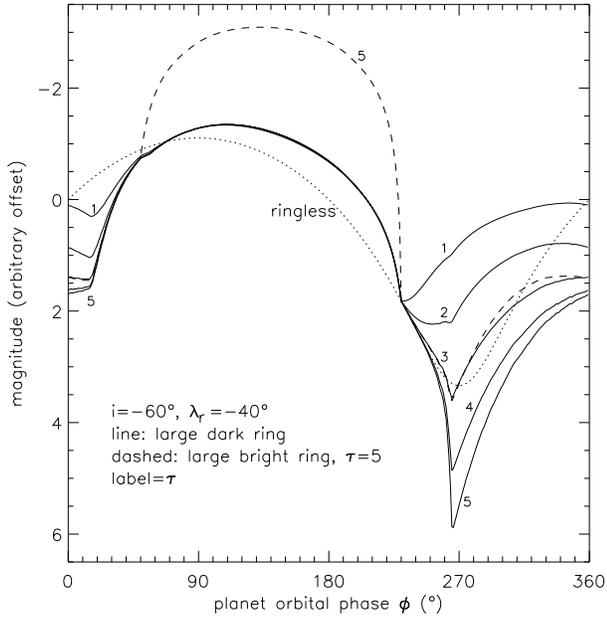}
   \caption{Ringed planet photometry versus $\phi$ for a large and dark ring with albedo $\varpi_0=0.05$ and different $\tau$ values (Table \ref{values}  and
   Fig. \ref{ap34_k100_rp30_ar5_tau400_ri33_ro81_ir40_lambdar-40_i-60_step1_image_along_orbit}). This dark ring hides the planet
   with a maximum around $\phi\approx265^{\circ}$ and makes the system fainter than a ringless planet for about one half orbit.
   A curve from Fig. \ref{anneau_saturnien_et_grand} for a planet with a large, bright and thick ring ($\varpi_0=0.7$ and $\tau=5$) is shown for comparison. All curves are normalized to the flux of
   a ringless planet seen at half phase.}
   \label{anneau_saturnien_et_grand_et_sombre}
\end{figure}

When the ring is thick and has a low albedo comparable to the albedo of asteroids (\cite{fernandez_et_al2003}), the transmitted
or reflected light contribution from the ring decreases. Moreover when the ring hides the planet, the brightness of the system decreases
by a factor of 15 in our example, making the system fainter than a ringless planet
for about the half of the orbit (Fig. \ref{anneau_saturnien_et_grand_et_sombre}).
As pointed out in Sect. \ref{section planet_no_ring}, from the operational point of view this means that a given star
must be monitored  for a time of the order of the orbital period.

We consider that such a strong and quite long extinction occurring at phase angles $\alpha>90^{\circ}$
constitutes another specific signature of a ring around an extrasolar planet. It was qualitatively
discussed by Schneider (in \cite{desmarais_et_al2002a}).


\subsubsection{ A ringed Earth-like planet }
\label{ringed_earth}
\begin{figure}
   \centering
   \includegraphics[width=8.75cm]{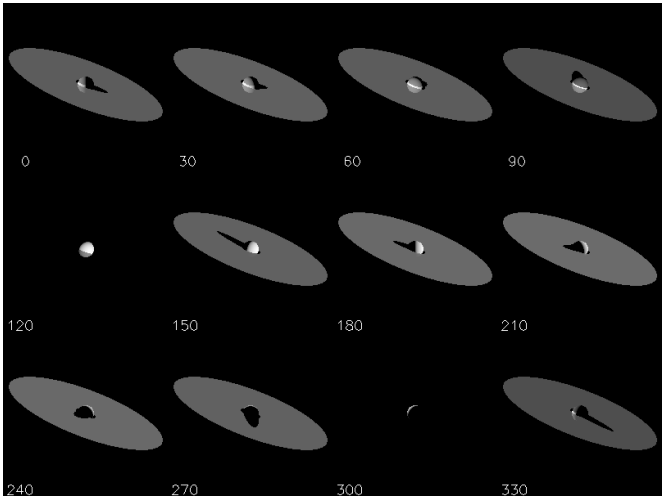}
   \caption{Ringed Earth-like planet for different orbital positions $\phi$ (labels). The ring outer radius is $r_o=10\ r_p$. Here, $\lambda_r=30^{\circ}$,
   $\varpi_0=0.05$ and $\tau=0.5$ (Table \ref{values}). Note the ring shadow on the planet, and also the planet seen through the ring,  at $\phi=90^{\circ}$ for instance.
   At the equinoxes $\phi=120^{\circ}$ and $300^{\circ}$,  part of the planet is obscured by the unilluminated ring ($\mu_0=0$ for the ring). Light curves are given
   in Fig. \ref{ringed_earth-like_planet_lambdar30}.
   Inclination $i=-60^{\circ}$. Image scale = radiance$^{0.33}$.}
   \label{ap30_k100_rp10_ar5_tau50_ri13_ro100_ir23_lambdar30_i-60_step30_image_along_orbit}
\end{figure}
\begin{figure}
   \centering
   \includegraphics[width=8.75cm]{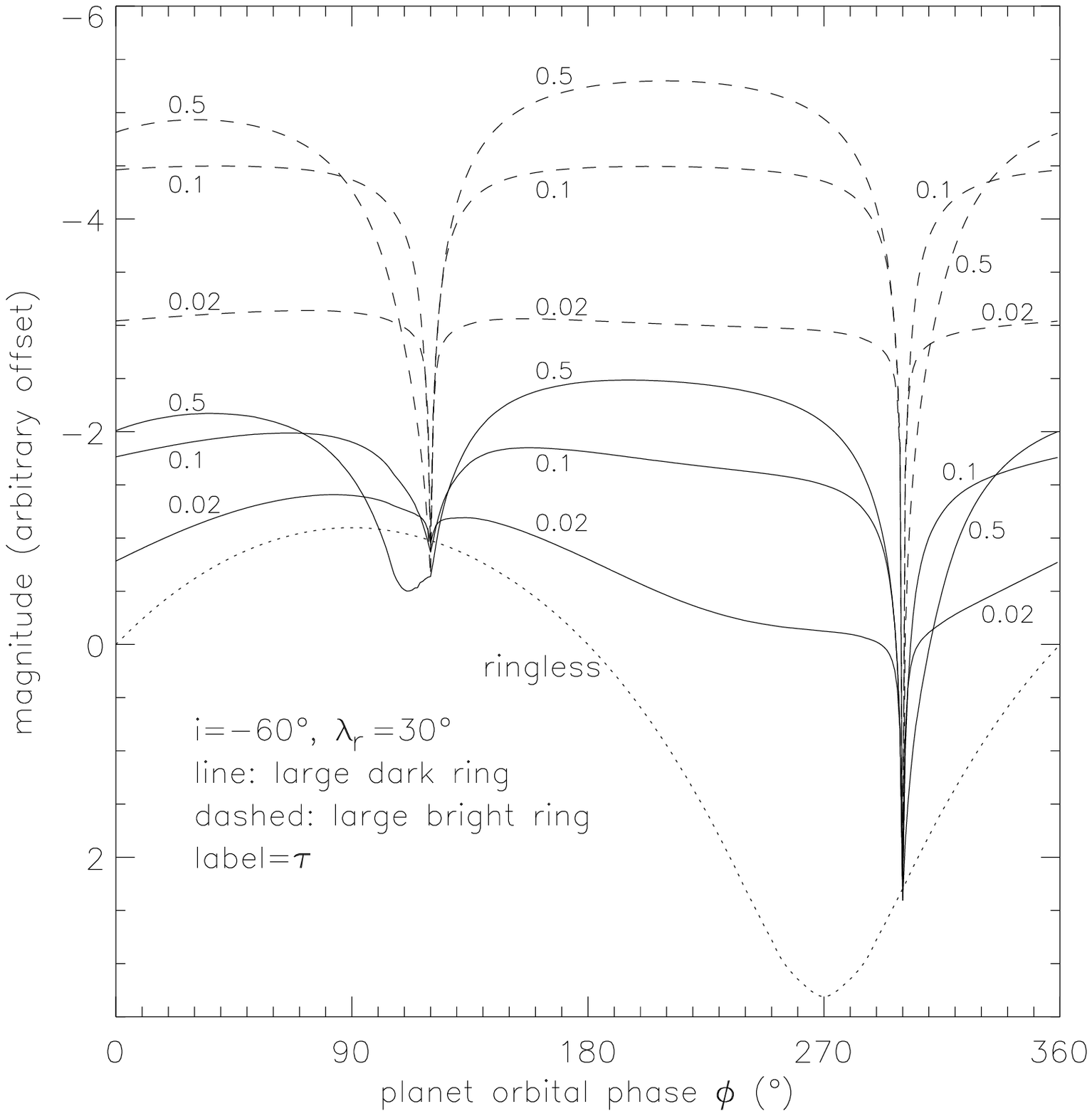}
   \caption{Ringed Earth-like planet photometry versus $\phi$ for a large ring with albedo $\varpi_0=0.7$ and 0.05. (Table \ref{values}
   and Fig. \ref{ap30_k100_rp10_ar5_tau50_ri13_ro100_ir23_lambdar30_i-60_step30_image_along_orbit}).
   All curves are normalized to the flux of a ringless planet seen at half phase.}
   \label{ringed_earth-like_planet_lambdar30}
\end{figure}
We mentioned in Sect. \ref{section_ring} that some authors have proposed that the Earth may have been surrounded by temporary rings,
created after the encounter with comets or asteroids.
We study here two kinds of rings: i) a dark dusty ring, considering that at only $1\ AU$ from the Sun, a ring of ice would not survive for a long time, and ii)
a Saturn-like bright ring, assuming an exo-Earth at a larger distance from its star ($>1.5\ AU$), where ice particles would survive. Here we consider a large
ring of outer radius $r_0=10\ r_p$ (Table \ref{values}). The brightness of a planet with such a large ring can be much higher than that of a ringless planet.
In our example (Fig.
\ref{ap30_k100_rp10_ar5_tau50_ri13_ro100_ir23_lambdar30_i-60_step30_image_along_orbit} and \ref{ringed_earth-like_planet_lambdar30}), the ringed planet
is typically 5 magnitudes, i.e. $100\times$ brighter than a ringless planet. But it is known that the flux of the reflected light from the planet $F_p$ divided
by the stellar flux $F_s$ is given by (\cite{schneider01}, 2002)
\begin{eqnarray}
\label{flux_ratio}
{F_p\over F_s}={A_p\over4}\bigg({r_p\over a}\bigg)^2 f(\phi)
\end{eqnarray}
where $a$ is the distance of the planet to its star and $f(\phi)$ a phase factor. Therefore, a $100\times$ overestimated flux ratio would give
a planet radius 10 times too large, which would be incorrectly estimated as $r_p\approx r_o\sqrt|\mu|$. If the mass of the planet is known (by radial velocity for instance),
the overestimated planet radius would lead
to a planet density underestimated by a factor of 1000. Thus a big planet of very low mass, i.e. with a very low apparent density, could be a signature of a ring.

Moreover, the planet radius can in principle be deduced from its infrared thermal flux $F_{p,IR}$, assuming the planet temperature has been inferred from the matching of a
Planck function to the observed thermal spectrum (\cite{schneider01}, 2002),
\begin{eqnarray}
\label{irflux}
F_{p,IR}= 4\pi\sigma_s\ T_p^4\ r_p^2,
\end{eqnarray}
where $\sigma_s$ is the Stefan constant and $T_p$ the system temperature. If planet and ring have the same temperature, then the deduced radius will be overestimated
as in the visible. But if the planet has its own thermal emission, like Saturn (\cite{allen1983}), the planet is much warmer than the ring which consequently does
not contribute significantly to the thermal flux. In that case, the planet
radius deduced from the IR flux would be smaller that the value found at visible wavelengths. The disagreement between the planet radii at visible and
infrared wavelengths could thus be another signature of a ring.


\section{Conclusions}
This work demonstrates that a ring around an extrasolar planet significantly affects the reflected light curve of an extrasolar
planet during its orbital motion. A ring could thus be detected, although both planet and ring would obviously remain unresolved. This may
be achieved simply by photometric monitoring of the planet reflected light. We identified the following signatures, which would require only
moderate photometry accuracy to be observed (around $3\sigma\approx0.5$ mag):

i) {\it Light curve dichotomy, with strong slope changes at the equinoxes}, due to the ring being alternately seen in reflection and in transmission.

ii) {\it Light curve dependence on wavelength} in dual-band photometry (methane band and continuum for instance), or
{\it spectral variations} if spectroscopy is possible.

iii) The {\it $\phi$-shift} of the light curve extrema, due to the longitude of the ring obliquity.

iv) {\it Temporary extinction of the planet} during the orbital motion, due to the rising shadow of the ring on the planet.

v) {\it High brightness in the reflected light}, leading to abnormally large planet radius, or/and abnormally low mass density.

vi) {\it Disagreement between the planet radii} measured from reflected light and thermal infrared emission.

Although future space missions studies concentrate mainly on infrared instruments for technical and scientific
reasons, this work shows the additional interest of shorter wavelengths (visible band) for extrasolar planet characterization.

\begin{acknowledgements}
      The authors are grateful to J. Berthier, M. Fairbairn, T. Guillot, D. Hestroffer and J. Tatum for helpful discussions about planets
      and radiometry.
\end{acknowledgements}

\end{document}